\documentclass[12pt, prd, showpacs]{revtex4}
\usepackage{amssymb}

\input{tcilatex}

\begin{document}

\title{Extrinsically flat static spacetimes }
\author{O. B. Zaslavskii}
\affiliation{Department of Mechanics and Mathematics, Kharkov V.N.Karazin National
University, \\
Svoboda Square 4, Kharkov 61077, Ukraine}
\email{ozaslav@kharkov.ua}

\begin{abstract}
We consider static spacetimes whose spatial part admits foliations with the
extrinsic curvature tensor $K_{ab}=0$. There are two complementary cases
when the gradient of the lapse function points 1) to the direction of
foliation or 2) orthogonally to it. Case 1) gives generalization of metrics
like Bertotti-Robinson or Nariai. In case 2) the matter source violates the
null energy condition at least on the part of the manifold, having in this
sense phantom nature. We also demonstrate that for the manifolds under
discussion the horizon can be naked in the sense that certain Weyl
components diverge in the free-falling frame although the Kretschmann scalar
is finite. The Petrov type is D or O. Explicit solutions for (i) the linear
anisotropic equation of state, (ii) Chaplygin gas and (iii) uniform energy
density are found.
\end{abstract}

\maketitle

\section{Introduction}

In the past decade, spacetimes which represent direct product of the
two-dimensional Lorentzian manifold (typically, adS, dS or Rindler) with a
sphere of a constant radius attract special interest. This happens, in
particular, in the context of string theory and adS/CFT correspondence \cite%
{1}, \cite{2}. In black hole physics they appear as a result of some
limiting transition from the non-extremal state to the extremal one \cite%
{prl}, \cite{extr}. Such a procedure can be applied to a generic static
black hole spacetime \cite{hm} including, for example, C-metric \cite{jose}.
Similar geometries occur also in non-linear electrodynamics \cite{jurek},
problems with string dust \cite{dad}, higher-dimensional \cite{vanzo} - \cite%
{flux} or 2+1 dimensional spacetimes \cite{c4} - \cite{jz}, dilaton-axion
gravity \cite{gc}, etc.

Formally, such spacetimes can be obtained if in the generic form of the
static spherically-symmetric line element one puts the value of the areal
radius $r=r_{0}=const$. The famous examples are Bertotti-Robinson (BR) \cite%
{br} and Nariai \cite{nar} spacetimes. These geometries were discussed
recently for constructing composite regular spacetimes \cite{z06} which are
glued with the metric of a black hole, wormhole or gravastar \cite{maz}
outside. Meanwhile, a natural question arises, whether one can enumerate all
possible metrics of this structure without the special assumption about the
symmetry of the spatial geometry. Mathematically, this means that one should
put for the extrinsic curvature tensor $K_{ab}=0$ for some form of embedding
the two-dimensional surface which is not necessarily a sphere (see details
below). Metrics of such a kind (called extrinsically flat) are encountered,
for instance, in wormhole physics as tube-like geometries \cite{visgen}, 
\cite{dav47}, \cite{antitr}. It has been also pointed out that consistent
boundary conditions for supergravity require vanishing extrinsic curvature
of the boundary \cite{nv}.

There is one more aspect connected with spacetimes under discussion that
deserves attention by itself. For the BR spacetime the longitudinal pressure 
$p_{\parallel }=-\rho $, where $\rho $ is the energy density. Thus, the
null-energy condition (NEC)\ is satisfied on the verge. We will see below
that in general there are also extrinsically flat metrics with $p_{\parallel
}+\rho <0$. This is nothing else than so-called phantom energy. Recently, it
focused much attention because of qualitatively new features in cosmology
where it drives acceleration of Universe \cite{reiss}, \cite{per}. However,
it is of interest also in wormhole physics where violation of NEC is
necessary condition for the existence of traversable wormholes \cite{mt}, 
\cite{vis}. Recently, explicit examples of wormholes supported of such a
kind of matter were analyzed numerically and analytically for anisotropic
phantom energy \cite{sushkov} - \cite{pz}. In this sense, the present work
extends the class of such objects and suggests some exact static solutions
with anisotropic phantom energy.

\section{2+1+1 decomposition of Einstein equations}

In the Gauss normal coordinates any static metric can be written in the form

\begin{equation}
ds^{2}=-dt^{2}N^{2}+dn^{2}+\gamma _{ab}dx^{a}dx^{b}{,}  \label{m}
\end{equation}%
where $x^{1}=n$, $a=2,3$. We will exploit the convenient representation of
the curvature tensor based on 2+1+1 decomposition of the metric \cite{israel}%
, \cite{med} that extensively uses the notion the extrinsic curvature tensor 
$K_{ab}$ for the surface $t=const$, $n=\nolinebreak const$ embedded in the
outer three-space. For our metric (\ref{m}) 
\begin{equation}
K_{ab}=-\frac{1}{2}\frac{\partial \gamma _{ab}}{\partial n}\text{.}
\label{kab}
\end{equation}%
In what follows we will not list the complete formulas for the Einstein
tensor (a reader can found them in \cite{israel}, \cite{med}) but take into
account at once their simplified form in which we put $K_{ab}=0$ from the
very beginning: 
\begin{equation}
^{(3+1)}G_{ab}=-\frac{N_{;a;b}}{N}+\gamma _{ab}(\frac{\Delta _{2}N}{N}+\frac{%
N^{\prime \prime }}{N})\text{,}  \label{ab}
\end{equation}%
\begin{equation}
G_{na}=-\frac{\partial _{n}N_{;a}}{N}\text{,}  \label{na}
\end{equation}%
\begin{equation}
G_{nn}=-\frac{1}{2}R_{\parallel }+\frac{\Delta _{2}N}{N},  \label{nn}
\end{equation}%
\begin{equation}
\frac{G_{00}}{N^{2}}=\frac{1}{2}R_{\parallel }\text{,}  \label{00}
\end{equation}%
prime denotes differentiation with respect to $n$, quantities $N_{;b;d}$
correspond to covariant derivatives with respect to the two-metric $\gamma
_{ab}$, $\Delta _{2}$ is the corresponding Laplacian, $R_{\parallel }$
represents the two-dimensional Ricci scalar for the surface $t=const$, $%
n=const$, $R_{\perp }=-2N^{-1}\frac{\partial ^{2}N}{\partial n^{2}}$is the
similar quantity for the $n-t$ subspace. We also suppose that our
stress-energy tensor has the form

\begin{equation}
T_{\mu }^{\nu }=diag(-\rho ,p_{\parallel },p_{\perp },p_{\perp })\text{,}
\label{tmn}
\end{equation}%
where $p_{\parallel }$ and $p_{\perp }$ are the longitudinal and transversal
pressure, respectively, and $\rho $ is the energy density.

Let $K_{ab}=0$ in some region, so according to (\ref{kab}) the
two-dimensional metric $\gamma _{ab}=\gamma _{ab}(x^{2},x^{3})$ does not
depend on $n$. It follows from eq. (\ref{na}) that $\frac{\partial ^{2}N}{%
\partial n\partial x^{a}}=0$, whence 
\begin{equation}
N=N_{1}(n)+N_{2}(x^{a}).  \label{nxa}
\end{equation}%
For simplicity, we consider separately in detail two cases when only either
first or second term in (\ref{nxa}) is present. More general case, when both
terms are non-vanishing, is discussed briefly in Sec. X.

\section{Class 1.}

Let $N=N_{1}(n)$, $N_{2}(x^{a})=0$. Then it follows from eqs. (\ref{ab}) - (%
\ref{00}) that%
\begin{equation}
\rho +p_{\parallel }=0\text{,}  \label{rop}
\end{equation}%
\begin{equation}
\frac{1}{2}R_{\parallel }=8\pi \rho \text{, }  \label{pr}
\end{equation}%
\begin{equation}
-\frac{R_{\perp }}{2}=\frac{1}{N}\frac{d^{2}N}{dn^{2}}=8\pi p_{\perp }\text{.%
}  \label{Nn}
\end{equation}%
In the case under discussion the transversal pressure $p_{\perp }$ depends
on $n$ only, while the longitudinal pressure $p_{\parallel }$ and energy
density depend on $x^{2}$, $x^{3}$ only. In the particular case $p_{\perp
}=const$ the situation simplifies further. Depending on the sign of $%
p_{\perp }$, we have three different subcases (up to the freedom in the
normalization of $N$).

A) $\lambda \equiv 8\kappa p_{\perp }=\kappa ^{2}>0$. Then a) $N=\sinh
\kappa n$, b) $N=\cosh \kappa n$, c) $N=\exp (\kappa n)$. Thus, we obtain
generalization of the BR \cite{br} spacetime, the $t-n$ part of the metric
being the two-dimensional (2D) adS spacetime.

B) $\lambda =-\kappa ^{2}<0$. Then $N=\sin \kappa n$, we have generalization
of the Nariai spacetime \cite{nar}, the $t-n$ part of the metric being the
2D dS spacetime.

C) $p_{\perp }=0$. Then a) $N=n$ (2D Rindler metric), b) $N=1$ (2D Minkowski
metric).

All three subcases A) - C) represent direct generalization of the metric%
\begin{equation}
ds^{2}=-N^{2}dt^{2}+dn^{2}+r^{2}(d\theta ^{2}+\sin ^{2}\theta d\phi ^{2})
\end{equation}%
with $r=r_{0}=const$ to the case with an arbitrary $\gamma
_{ab}(x^{2},x^{3}) $.

The positivity of the energy density entails from (\ref{rop}), (\ref{pr})
that $R_{\parallel }>0$ and, thus, the 2D manifold $x^{2}$, $x^{3}$ is
convex.

\section{Class 2}

Let $N=N(x^{2},x^{3})$. Then it follows from Einstein equations (\ref{ab}) -
(\ref{00}) that eq. (\ref{pr}) holds and, apart from this,%
\begin{equation}
\frac{\Delta _{2}N}{N}-\frac{1}{2}R_{\parallel }=8\pi p_{\parallel }\text{,}
\label{22}
\end{equation}%
\begin{equation}
-\frac{N_{;a;b}}{N}+\gamma _{ab}\frac{\Delta _{2}N}{N}=8\pi p_{\perp }\gamma
_{ab}\text{.}  \label{23}
\end{equation}

From (\ref{pr}) and (\ref{22}) - (\ref{23}) we have 
\begin{equation}
\frac{\Delta _{2}N}{N}=8\pi (\rho +p_{\parallel })=16\pi p_{\perp }\text{,}
\label{ppn}
\end{equation}%
whence%
\begin{equation}
\rho +p_{\parallel }-2p_{\perp }=0\text{.}  \label{pp}
\end{equation}%
In the isotropic case $p_{\parallel }=p_{\perp }=p$ we obtain the perfect
stiff fluid but, in general, we assume that pressures $p_{\parallel }$ and $%
p_{\perp }$ do not coincide. By substitution of (\ref{ppn}) into (\ref{23})
we obtain

\begin{equation}
\frac{N_{;a;b}}{N}=\lambda \gamma _{ab}\text{, }  \label{nab}
\end{equation}%
\begin{equation}
\lambda \equiv 8\pi p_{\perp }=\frac{\Delta _{2}N}{2N}\text{.}  \label{pl}
\end{equation}%
It is worth noting that equations (\ref{nab}), (\ref{pl}) entail that the
regularity of the four-dimensional (4D) geometry is encoded in the
finiteness of the quantity $\lambda $. Indeed, the Kretschmann scalar%
\begin{equation}
Kr=R_{\alpha \beta \gamma \delta }R^{\alpha \beta \gamma \delta
}=P_{ijkl}P^{ijkl}+4\frac{N_{i\mid j}N^{i\mid j}}{N^{2}}\text{,}  \label{inv}
\end{equation}%
$R_{\alpha \beta \gamma \delta }$ is the 3+1 Riemann tensor, $P_{ijkl}$ is
the curvature tensor of the hypersurface $t=const$, $N_{i\mid j}$ denotes
covariant derivatives with respect to the three-metric on it. Now only terms
with $i,j=a,b$ survive in the second term in (\ref{inv}). Taking into
account (\ref{nab}) we see that $Kr=P_{ijkl}P^{jkl}+8\lambda ^{2}$. As the
first term in $Kr$ refers to the three-dimensional positively defined
metric, it is positive by itself, so we have two positive (or, at least,
non-negative) terms, each of them should be finite if we want the finiteness
of $Kr$.

It follows from eq. (\ref{nab}) that the vector $\xi _{a}\equiv N_{;a}$
satisfies the conformal Killing equation%
\begin{equation}
\xi _{(a;b)}=\frac{1}{2}\xi _{;a}^{a}\gamma _{ab}\text{.}
\end{equation}%
If we define the new vector

\begin{equation}
\text{ }\eta _{a}\equiv \varepsilon _{ac}\xi ^{c}\text{,}  \label{nx}
\end{equation}%
where $\varepsilon _{ac}=\sqrt{\gamma }e_{ac}$, $e_{ac}$ is the absolutely
antisymmetric symbol ($e_{01}=1$), we obtain

\begin{equation}
\eta _{a;b}+\eta _{b;a}=0\text{.}
\end{equation}%
Thus, the vector $\eta _{a}$ is the Killing one. It is seen from (\ref{nx})
that $\eta _{a}\xi ^{a}=0$. Thus, the Killing vector is pointed orthogonally
to the gradient of $N$ and, therefore, along the lines of the constant $N$.
This prompts us to choose the convenient coordinate system in such a way
that the lines of constant $N$ be functions of $x$ alone, while the Killing
vector $\eta =\frac{\partial }{\partial \phi }.$Then, using the proper
length as a coordinate, we have for the metric of $x-\phi $ manifold 
\begin{equation}
ds_{2}^{2}=dx^{2}+r^{2}(x)d\phi ^{2}\text{,}  \label{mb}
\end{equation}%
so that our 4D metric can be written in the explicitly axially-symmetric
form:%
\begin{equation}
ds^{2}=-N^{2}(x)dt^{2}+dn^{2}+dx^{2}+r^{2}(x)d\phi ^{2}\text{.}  \label{4d}
\end{equation}%
Thus, the metric turns out to be homogeneous in $n$ direction and is
symmetric under rotation in $\phi $ direction.

We suppose that the condition $\rho >0$ is satisfied, so that according to (%
\ref{pr}) $R_{\parallel }>0$ everywhere as well as it happens to class 1
metrics. Thus, the 2D $x-\phi $ manifold is convex everywhere. We discuss
separately two cases: the manifold is (i) compact and simply connected, (ii)
non-compact.

\section{Compact 2D manifold: phantom nature of source and properties of
horizon}

The variable $x$ changes in the finite range $0\leq x\leq x_{1}$. It is also
convenient to normalize $\phi $ as usual, $0\leq \phi \leq 2\pi $. The
function $r$ should ensure the regularity of the metric near the north and
south poles, whence 
\begin{equation}
r=x\text{, }x\rightarrow 0\text{ and }r=x_{1}-x\text{ at }x\rightarrow x_{1},
\label{01}
\end{equation}%
where for definiteness we assume that $r>0$. If $r$ has additional zeros at $%
0<x<x_{1}$, this would simply mean that we are faced with the case when our
construction is obtained by gluing chain of ball-like manifolds, with the
south pole of the n-th copy coinciding with the north pole of the n+1 north
one. For simplicity, we discuss properties of a single manifold only and
assume that the function $r(x)$ has a definite sign inside the interval ($%
0,x_{1}$).

The scalar curvature for the metric (\ref{mb}) is equal to 
\begin{equation}
R_{\parallel }=-2\frac{r^{\prime \prime }}{r}\text{,}  \label{rb}
\end{equation}%
prime denotes differentiation with respect to $x$. It follows from (\ref{pr}%
) and (\ref{rb}) that%
\begin{equation}
\frac{r^{\prime \prime }}{r}=-8\pi \rho \text{.}  \label{rro}
\end{equation}

The fact that $R_{\parallel }>0$ restricts the behavior of $r(x)$ in the
following way. As we chose $r>0$ inside the interval $0<x<x_{1}$, we have $%
r^{\prime \prime }<0$ there. As $r=0$ at $x=0$ and $x=x_{1}$, its derivative
changes sign at some $x_{0}$ where $r^{\prime }(x_{0})=0$. Because of $%
r^{\prime \prime }<0$ the derivative $r^{\prime }$ may vanish only in one
point, so there exists only one root of equation $r^{\prime }(x_{0})=0$.

One can see that $x\phi $ equation from the set (\ref{nab}) holds
identically for the metric (\ref{mb}), while $\phi \phi $ and $xx$ ones give
us after simple manipulations that%
\begin{equation}
N=2\pi c\int^{x}dxr=c(A+A_{0}),c=const\text{, }A_{0}=const\text{.}  \label{n}
\end{equation}%
\begin{equation}
\lambda =\frac{r^{\prime }}{\int_{0}^{x}dxr+N_{0}}\text{, }N_{0}=\frac{A_{0}%
}{2\pi }\text{,}  \label{l}
\end{equation}%
now prime means differentiation with respect to $x$. The quantity $A>0$
represents the area of the region between the north pole and the line $%
N(x)=const$. In what follows we normalize the lapse function in such a way
that $c=\frac{1}{2\pi }$.

If the function $r(x)$ \ is known, so that the 2D metric $\gamma _{ab}$ is
fixed, one can recover from eqs. (\ref{pr}), (\ref{pp}), (\ref{n}) and (\ref%
{l}) all the rest of information: the lapse function $N(x)$, both pressures $%
p_{\parallel }$ and $p_{\perp }$ and the energy density $\rho $.
Alternatively, one can define $T_{\mu }^{\nu }$ (\ref{tmn}) with the
equation of state, say, $p_{\perp }=p_{\perp }(\rho )$ supplemented by eq. (%
\ref{pp}). Then one has to solve one equation in which $p_{\perp }$ is
defined according to eq. (\ref{l}) and $\rho $ is defined according to eqs. (%
\ref{pr}), (\ref{rb}). Afterwards, the lapse function can be found from (\ref%
{n}). Eq. (\ref{l}) can be rewritten in the form of the second order
differential equation%
\begin{equation}
N^{\prime \prime }-\lambda N=0\text{,}  \label{z}
\end{equation}%
where 
\begin{equation}
N=\int_{0}^{x}dxr+N_{0},N_{0}=N(0)\text{, }r=N^{\prime }\text{,}  \label{zbx}
\end{equation}%
This equation is supplemented by the conditions $N"(0)=1$ and $N"(x_{1})=-1$
as it follows from (\ref{01})$.$Hence, $N(0)=\lambda ^{-1}(0)$ and $%
N(x_{1})=-\lambda ^{-1}(x_{1})$.

One can rescale the coordinate according to $x=x(y)$ and obtain in (\ref{mb}%
) $ds_{2}^{2}=\chi ^{2}(dy^{2}+C^{2}d\phi ^{2})$, where $\chi =\frac{dx}{dy}$%
, $C(y)=\frac{r(x(y))}{\chi (y)}$. Choosing $y$ in such a way that $\frac{dy%
}{\sin \frac{y}{r_{0}}}=\frac{r_{0}dx}{r}$, $\chi \equiv e^{\omega }$, $y$ $%
\equiv r_{0}\theta $ ($r_{0}$ is a constant of dimensionality of length), we
obtain 
\begin{equation}
ds_{2}^{2}=r_{0}^{2}e^{2\omega (\theta )}(d\theta ^{2}+\sin ^{2}\theta d\phi
^{2})\text{.}  \label{y}
\end{equation}%
It is written in the explicitly axially symmetric form and is conformal to
the metric on the unit sphere. Then expressions (\ref{n}), (\ref{l}) for $N$
and $\lambda $ can be rewritten as%
\begin{equation}
N=r_{0}^{2}\int d\theta \exp (2\omega )\sin \theta \text{,}
\end{equation}%
\begin{equation}
\lambda =\frac{\cos \theta +\omega ^{\prime }\sin \theta }{N}\text{.}
\end{equation}

Let us now discuss some properties of the solution. If we want regularity,
either a) $N_{0}>0$ (then $N>0$ vanishes nowhere) or b) we must choose the
constant $N_{0}$ in such a way that $N_{0}=-\int_{0}^{x_{0}}rdx$, 
\begin{equation}
N=\int_{x_{0}}^{x}dxr\text{,}  \label{zx}
\end{equation}
then the zero of the denominator in $\lambda =\frac{r^{\prime }}{N}$ is
compensated by the zero of the numerator. In doing so, $N(x_{0})=0$, so we
have a Killing horizon at $x_{0}$.

In case a) it follows from (\ref{l}) that $\lambda >0$ at $x<x_{0}$ (where
the function $r(x)$ is monotonically increasing) and $\lambda <0$ at $%
x>x_{0} $ (where $r(x)$ is monotonically decreasing). It follows from (\ref%
{pp}) that negative $\lambda $ implies that $\rho +p_{\parallel }<0$ and,
thus, the NEC (null energy condition) is violated, so the matter is phantom
on the part $x>x_{0}$ of the manifold.

Consider case b)$.$Then, at $x<x_{0}$, $r^{\prime }>0$ but the denominator
in (\ref{l}) is negative. For $x>x_{0}$ the situation is reverse. At the
point $x_{0}$, exploiting the Lhopital's rule, we obtain that on the horizon 
\begin{equation}
\lambda =\frac{r^{\prime \prime }(x_{0})}{r(x_{0})}<0.  \label{horp}
\end{equation}%
Thus, the matter is phantom everywhere.

Taking into account (\ref{rro}) we see that at the horizon itself 
\begin{equation}
p_{\perp }=-\rho \text{, }p_{\parallel }=-3\rho \text{.}  \label{phor}
\end{equation}%
.

The typical asymptotic behavior near the horizon 
\begin{equation}
N=\kappa _{H}n+\frac{\kappa _{2}}{3!}(x^{2},x^{3})n^{3}+...\text{,}
\label{N}
\end{equation}%
where $\kappa _{H}=const$ is the surface gravity \cite{med} does not hold
literally since $N$ does not depend on $n$ at all but its counterpart is
valid if the variable $n$ is replaced by $x$:

\begin{equation}
N=r(x_{0})(x-x_{0})+\frac{r^{\prime \prime }(x_{0})}{6}(x-x_{0})^{3}+...%
\text{, }\kappa =r(x_{0})\text{,}  \label{X}
\end{equation}%
it is assumed that the lapse function can be expanded in the Taylor series.
In a similar way, the equality $G_{n}^{n}-G_{0}^{0}=0$ typical of the
non-extremal regular horizon \cite{med} is replaced now by $%
G_{x}^{x}-G_{0}^{0}=8\pi (p_{\perp }+\rho )=0$ on the horizon.

It is instructive to consider the motion of particles along geodesics on the
surface $n=const$, when the horizon is present. Introducing the
four-velocity of an observer $u^{\mu }=\frac{dx^{\mu }}{d\tau }$ we have

$E=-u_{0}$, $u^{0}=\frac{E}{N^{2}}$ and it follows from the condition $%
u^{\mu }u_{\mu }=-1$ for time-like geodesics that%
\begin{equation}
\left( u^{x}\right) ^{2}-\frac{E^{2}}{N^{2}}=-1\text{,}
\end{equation}%
whence%
\begin{equation}
u^{x}=\frac{1}{\left\vert N\right\vert }\sqrt{E^{2}-N^{2}}\text{,}
\end{equation}%
so that%
\begin{equation}
u^{\mu }=(\frac{\cosh s}{\left\vert N\right\vert },0,\sinh s,0)\text{, }%
\cosh s=\frac{E}{\left\vert N\right\vert }\text{.}  \label{u}
\end{equation}%
Supposing that a particle moves with increasing $x$, we choose $u^{x}>0$. It
follows from (\ref{l}) that 
\begin{equation}
\dot{r}=\frac{dr}{d\tau }=-\lambda \sqrt{E^{2}-N^{2}}\text{,}
\end{equation}%
where we took into account that according to (\ref{zx}) $N<0$ for $x<x_{0}$.
As $\lambda <0$, $\dot{r}>0$ (except, possibly, the horizon itself where $%
\rho =\lambda =0$ is allowed - see next section). Thus, a particle crosses
the horizon with increasing the radial distance. This means that the
coordinate frame (\ref{4d}) that is suitable in the region $r\leq r(x_{0})$
containing the origin, does not cover the total manifold and one is forced
to introduce Kruskal-like coordinates in a standard way. Actually, the
situation is quite similar to what happens in the case of de Sitter metric
and, thus, we have a cosmological horizon, the topology of ($t,r,\phi $)
submanifold being RxS$^{2}$.

\section{Explicit examples. Compact 2D manifolds}

\subsection{Linear equation of state}

We will see in this section that, if the equation of state is known, the
above formulas may admit, in principle, simple explicit solutions. We
restrict ourselves by a physically important case of the linear connection
between the pressure and density. In this section we discuss compact 2D
manifolds. Let 
\begin{equation}
p_{\perp }=-\alpha \rho .  \label{ar}
\end{equation}%
As $p_{\perp }$ cannot be positive everywhere (see the previous section), we
must choose $\alpha >0$. Then $p_{\perp }<0$ everywhere if $\rho >0$ and,
thus, there is a horizon. Then it follows from (\ref{pp}) that $p_{\parallel
}=-(1+2\alpha )\rho $ and $p_{\parallel }+\rho =-2\alpha \rho <0$. Using
formulas (\ref{z}), (\ref{pr}) we find the equation%
\begin{equation}
\alpha \frac{r^{\prime \prime }}{r}=\frac{N^{\prime \prime }}{N}\text{,}
\end{equation}%
whence, taking into account that $r=N^{\prime }$, 
\begin{equation}
\left\vert N^{\prime \prime }\right\vert =N_{1}\left\vert N\right\vert
^{\gamma }\text{, }\gamma =\frac{1}{\alpha }>0\text{,}  \label{zz}
\end{equation}%
$N_{1}>0$ is a constant$.$It follows from (\ref{zz}) that in both regions $%
0\leq x\leq x_{0}$ and $x\geq x_{0}$%
\begin{equation}
Y^{\prime \prime }=-N_{1}Y^{\gamma }\text{, }Y=\left\vert N\right\vert \text{%
.}  \label{ny}
\end{equation}

Eq.(\ref{ny}) describes a motion of a particle with the unit mass in the
effective potential%
\begin{equation}
U(Y)=\frac{N_{1}}{\gamma +1}Y^{\gamma +1}\text{, }  \label{yz}
\end{equation}%
whence%
\begin{equation}
\frac{Y^{\prime 2}}{2}+U(Y)=\varepsilon \text{,}  \label{ez}
\end{equation}%
$\varepsilon $ is the effective "energy". Its value can be found by putting $%
x=x_{0}$, then $Y=0$ and $U=0$, $\varepsilon =\frac{r_{0}^{2}}{2}$, $%
r_{0}=r(x_{0})$.

It follows from (\ref{ez}) that%
\begin{equation}
N=\chi (y)\theta (x-x_{0})\text{, }y=r^{2}\text{, }\chi (y)=\left[ \frac{%
\left( y_{0}-y\right) (\gamma +1)}{2N_{1}}\right] ^{\frac{1}{\gamma +1}}%
\text{, }y_{0}=y(x_{0})=r_{0}^{2}\text{, }  \label{zb}
\end{equation}%
$\theta (x-x_{0})$ is the Heaviside step function. At $x_{0}$ $\chi =0$, $%
N=0 $. Hence, remembering that $N^{\prime }=r$, we find%
\begin{equation}
x=-2\int_{0}^{r}dr\frac{d\chi }{dy}\text{, }0\leq x\leq x_{0}\text{,}
\label{x1}
\end{equation}%
\begin{equation}
x=x_{0}+2\int_{r_{0}}^{r}dr\frac{d\chi }{dy}\text{, }x_{0}\leq x\leq x_{1}%
\text{,}  \label{x2}
\end{equation}%
where the signs are chosen in such a way that $r$ grows from zero at the
north pole to $r(x_{0})$ and then diminishes to zero at the south pole. In
general, these integrals are not given in terms of elementary functions.
However, one may use the quantity $r$ instead of $x$ to obtain an explicit
expression for the metric%
\begin{equation}
ds^{2}=-N^{2}dt^{2}+dn^{2}+dr^{2}f^{2}(r)+r^{2}d\phi ^{2}\text{,}  \label{f}
\end{equation}%
where $N^{2}=N^{2}(x(r))$, $f(r)=\mp 2\frac{d\chi }{dy}(y=r^{2})$ and $x$ is
given by eqs. (\ref{x1}), (\ref{x2}). Substituting explicit expressions for $%
\chi $ and taking into account (\ref{n}), (\ref{zbx}) we obtain%
\begin{equation}
N=const(r_{0}^{2}-r^{2})^{\delta }\text{, }\delta =\frac{\alpha }{1+\alpha }%
\text{,}  \label{Nr}
\end{equation}%
\begin{equation}
f=\frac{dx}{dr}=r_{0}^{2\beta }(r_{0}^{2}-r^{2})^{-\beta }\text{, }\beta =%
\frac{1}{1+\alpha }\text{, }  \label{fr}
\end{equation}%
$0\leq r\leq r_{0}$ and the constant $r_{0}$ in $f$ is chosen to respect the
condition $f(0)=1$ that follows from (\ref{01})$.$Thus, we obtained exact
solutions for equation of state (\ref{ar}).

If $\alpha =1=\gamma $, we obtain the metric on the sphere of a fixed
radius, $r=r_{0}\sin \frac{x}{r_{0}}$. Let now $\alpha \neq 1$. It follows
from (\ref{horp}) that the regularity of the horizon requires that $%
r^{\prime \prime }$ be finite there. Calculating it from (\ref{fr}), one
finds that at $r\rightarrow r_{0}$ 
\begin{equation}
r^{\prime \prime }\sim (r_{0}^{2}-r^{2})^{\sigma }\text{, }\sigma =\frac{%
1-\alpha }{1+\alpha }\text{, }x-x_{0}\sim (r_{0}^{2}-r^{2})^{\delta }\text{.}
\label{rn}
\end{equation}%
The horizon is regular if $\alpha <1$. Then $p_{\perp }+\rho =(1-\alpha
)\rho \geq 0$. Meanwhile, we remind that $p_{\parallel }+\rho =2p_{\perp }<0$%
. We see that the matter is phantom with respect to the pressure in $n$
direction but not in $x$ one$.$It follows from (\ref{rro}) that on the
horizon itself $\rho \rightarrow 0$, $r^{\prime \prime }\rightarrow 0$, $%
p_{\perp }\rightarrow 0$.

Near the horizon the lapse function behaves like 
\begin{equation}
N=\kappa _{H}(x-x_{0})+A(x-x_{0})^{q}+...\text{, }q=2+\frac{1}{\alpha }>3%
\text{.}  \label{N2}
\end{equation}%
where $A$ is a constant. This asymptotics is more general than (\ref{N}), (%
\ref{X}) listed in \cite{med} since it involves fractional powers. The
condition ensuring the finiteness of (\ref{inv}) is satisfied since $\frac{%
N^{\prime \prime }}{N}\sim (x-x_{0})^{d}$, $d=\frac{1-\alpha }{\alpha }>0$.

\subsection{Chaplygin equation of state}

Another type of equation of state that attracts much attention in recent
years is that of Chaplygin gas. It was exploited for modelling the dark
energy in cosmology \cite{ch1}, \cite{ch2} and quite recently was also used
in wormhole physics \cite{lobo2}. Now we apply it to the transversal
pressure:%
\begin{equation}
p_{\perp }=-\frac{A_{0}}{\rho }\text{, }A_{0}>0\text{,}
\end{equation}%
Taking into account (\ref{z}), (\ref{pr}), we obtain after integration: 
\begin{equation}
\left( N^{\prime \prime }\right) ^{2}=A^{2}N^{2}+A_{1}\text{,}  \label{z2}
\end{equation}%
where $A_{1}=const$, $A^{2}=(8\pi )^{2}A_{0}$. Remembering that on the
horizon $x=x_{0}$, $N=0$, $N^{\prime \prime }=0$, we immediately see that $%
A_{1}=0$. Then, it follows from (\ref{z2}) that $r=\frac{\sin \omega x}{%
\omega }$, $\omega =\sqrt{A}$ and, thus, our 2D manifold is a sphere. In
doing so, the energy density $\rho =\frac{\omega ^{2}}{8\pi }$ and $p_{\perp
}=-\frac{\omega ^{2}}{8\pi }$ turn out to be constants. The metric in ($%
t,x,\phi $) manifold is nothing else than de Sitter one, $\rho \,$plays the
role of the cosmological constant.

\subsection{Uniform density}

We may take the energy density $\rho \equiv \frac{\omega ^{2}}{8\pi }$ to be
a constant from the very beginning. Then (\ref{pr}) gives us again $r=\frac{%
\sin \omega x}{\omega }$, so that we have a sphere of a constant radius $%
r_{0}=\omega ^{-1}$. Using $\theta =\omega x$, we obtain%
\begin{equation}
N=a-\cos \theta \text{, }\lambda =\frac{r_{0}^{-2}\cos \theta }{a-\cos
\theta }\text{, }  \label{spp}
\end{equation}%
\begin{equation}
8\pi \rho =r_{0}^{-2}\text{, }8\pi p_{\perp }=r_{0}^{-2}\frac{\cos \theta }{%
a-\cos \theta }\text{, }8\pi p_{\parallel }=r_{0}^{-2}\frac{3\cos \theta -a}{%
a-\cos \theta }\text{, }
\end{equation}%
There are two possible subcases here.

a) $\left\vert a\right\vert >1$, let for definiteness $a>1$. Then $N>0$
everywhere, $\lambda $ changes the sign at $\theta =\frac{\pi }{2}$. For $%
\theta >\frac{\pi }{2}$ the matter is phantom with respect to $p_{\parallel
} $ ($\rho +p_{\parallel }$ $=2p_{\perp }<0$) but $p_{\perp }+\rho =\frac{1}{%
8\pi r_{0}^{2}}\frac{a}{a-\cos \theta }>0$. Thus, anisotropy is essential.

b) $\left\vert a\right\vert \leq 1$, $a=\cos \theta _{0}\,.$Then $N=0$ at $%
\theta =\theta _{0}$, $\theta =2\pi -\theta _{0}$, where $\lambda $ in
general diverges. Then the requirement of regularity allows only the value $%
\theta _{0}=\frac{\pi }{2}$, $a=0$, whence%
\begin{equation}
N=-\cos \theta \text{, }\rho =\frac{1}{8\pi r_{0}^{2}}\text{, }p_{\perp
}=-\rho \text{, }p_{\parallel }=-3p_{\perp }\text{.}  \label{st}
\end{equation}%
We saw that equation of state (\ref{st}) holds on the horizon, now it is
valid everywhere on the entire manifold, the 2+1 geometry is de Sitter one.

\section{Non-compact 2D manifold}

Now the variable $x$ changes in the range $0\leq x<\infty .$ One can easily
see that if $\rho >0$, the derivative $r^{\prime }$ cannot change its sign.
Indeed, let at some point $x_{0}$ $r^{\prime }(x_{0})=0$. As, by assumption, 
$r>0$, $r^{\prime \prime }<0$ this would mean that the function $r(x)$ would
inevitably vanish at some finite $x_{1}$ and we would return to the previous
case of a compact manifold in contradiction with the assumption. Thus, $%
r^{\prime }>0$ everywhere. In a similar way, this means that a non-compact
manifold should have the origin where $r=0$.

In turn, it follows from (\ref{l}) that a regular horizon is impossible now.
Indeed, on the horizon the denominator in $\lambda $ diverges whereas the
numerator $r^{\prime }$ does not vanish. As a result, it is incompatible
with the finiteness of $\lambda $. As $r^{\prime }>0$ and $N$ does not
vanish, the quantity $\lambda $ has the same sign everywhere. As $r^{\prime
}>0$, $r>0$, the integral $\int_{0}^{x}dxr>0$ and diverges at $x\rightarrow
\infty $. Thus, both the numerator and denominator in (\ref{l}) are
positive, so $\lambda >0$ everywhere and the matter is not phantom.

This entails that the equation of state (\ref{ar}) cannot be implemented.
Instead, one may try $p_{\perp }=\alpha \rho $ with $0<\alpha \leq 1$. Let $%
\alpha <1$. Now, the effective potential%
\begin{equation}
U=\frac{N_{1}}{\gamma -1}N^{1-\gamma }\text{, }\gamma =\frac{1}{\alpha }>1%
\text{, }N_{1}>0\text{.}  \label{uu}
\end{equation}%
It is seen that $U$ is a monotonically decreasing function, $U\rightarrow
\infty $ at $N\rightarrow 0$ and $U\rightarrow 0$ at $N\rightarrow \infty $.
It means that in eq. (\ref{ez}) $\varepsilon >0$, $Y=$ $N$ changes in the
semi-infinite interval from the turning point to infinity, $N_{0}\leq
N<\infty $, $0\leq r\leq r_{0}=\sqrt{2\varepsilon }$. Repeating previous
derivation, we find the metric in the form (\ref{f}) - (\ref{fr}) with 
\begin{equation}
\beta =\frac{1}{1-\alpha }>1
\end{equation}%
and 
\begin{equation}
\delta =-\frac{\alpha }{1-\alpha }\text{.}
\end{equation}

The explicit exact solutions under discussion coincide with those obtained
earlier in \cite{cf1}, \cite{cf2}. However, their counterparts (\ref{Nr}), (%
\ref{fr}) corresponding to the cosmological horizons are not contained in
these works where 2D manifolds were non-compact, while the solutions (\ref%
{Nr}), (\ref{fr}) imply that the 2D manifold is compact. (To avoid
confusion, it is worth noting that the statements in \cite{cf1}, \cite{cf2}
about finite proper volume actually refer to the space occupied by matter
whereas the total volume is infinite.) Asymptotically, as $r\rightarrow
r_{0} $, the metric of our 2D manifold looks like a cylinder%
\begin{equation}
ds_{2}^{2}=dx^{2}+r_{0}^{2}d\phi ^{2}
\end{equation}%
in agreement with \cite{cf2}, the proper length $x=\int drf\rightarrow
\infty $.

The case $\alpha =1$, according to (\ref{pp}), corresponds to stiff matter
with $p_{\parallel }=p_{\perp }=\rho $. It gives rise to the potential $%
U=-N_{1}\ln N$. Then%
\begin{equation}
N=\exp \left( \frac{r^{2}}{r_{1}^{2}}\right) \text{,}
\end{equation}%
\begin{equation}
f=\exp \left( \frac{r^{2}}{r_{1}^{2}}\right) \text{, }r_{1}=const\text{,}
\end{equation}%
now $0\leq r<\infty $. In this particular case the solution coincides with
that obtained in \cite{bar}.

For the non-compact case there is no sense to consider the Chaplygin
equation of state since it implies the negative pressure, whereas, as is
explained above, now $p_{\perp }>0$. The energy density cannot be uniform
since this would mean that $r\sim \sin \sqrt{8\pi \rho }x$ is bounded,
whereas our 2D manifold should be non-compact by assumption.

\section{Petrov type}

In this section, we determine the Petrov type of gravitation field which our
spacetimes belong to.

\subsection{Case 2}

We discuss firstly class 2. To this end, we proceed in a standard way. The
determination of the Petrov type is based on studying curvature invariants
constructed from so-called Weyl scalars obtained by the contraction of the
Weyl tensor with components of some complex null tetrad (see, e.g. \cite%
{kramer}). It is convenient to construct this tetrad from an usual
orthonormal frame $u^{\mu }$, $e^{\mu }$, $a^{\mu }$, $b^{\mu }$, where $%
u^{\mu }$ is a four-velocity of an observer, $e^{\mu }$ is a vector aligned
along the $n$-direction, $a^{\mu }$ and $b^{\mu }$ lie in the $x^{2}-x^{3}$
subspace. We define

\begin{equation}
l^{\mu }=\frac{u^{\mu }+e^{\mu }}{\sqrt{2}}\text{, }n^{\mu }=\frac{u^{\mu
}-e^{\mu }}{\sqrt{2}}\text{, }m^{\mu }=\frac{a^{\mu }+ib^{\mu }}{\sqrt{2}}%
\text{, }\bar{m}^{\mu }=\frac{a^{\mu }-ib^{\mu }}{\sqrt{2}}
\end{equation}%
(bar denotes complex conjugate)$.$ Now $l^{\mu }n_{\mu }=-1$, $m^{a}\bar{m}%
_{a}=1$, all other contractions vanish. We define the Weyl scalars according
to%
\begin{equation}
\psi _{0}=C_{s\beta \gamma \delta }l^{s}m^{\beta }l^{\gamma }m^{\delta }%
\text{, }\psi _{1}=C_{s\beta \gamma \delta }l^{s}m^{\beta }l^{\gamma
}n^{\delta }\text{, }\psi _{2}=-C_{s\beta \gamma \delta }l^{s}m^{\beta
}n^{\gamma }\bar{m}^{\delta }\text{,}  \label{psi02}
\end{equation}%
\begin{equation}
\psi _{3}=C_{s\beta \gamma \delta }l^{s}n^{\beta }\bar{m}^{\gamma }n^{\delta
}\text{, }\psi _{4}=C_{s\beta \gamma \delta }n^{s}\bar{m}^{\beta }n^{\gamma }%
\bar{m}^{\delta }\text{,}  \label{psi34}
\end{equation}%
$C_{s\beta \gamma \delta }$ is the Weyl tensor. One reservation is in order.
It was shown in \cite{vojta} that in the horizon limit the Petrov type may
be different in the static frame and the one attached to a free falling
observer. Correspondingly, we will distinct these both frames (not only on
the horizon). In the first case $u^{\mu }=(\left\vert N\right\vert
^{-1},0,0,0)$ and $e^{\mu }=(0,1,0,0)$, so that%
\begin{equation}
l^{\mu }=\frac{1}{\sqrt{2}}(\frac{1}{\left\vert N\right\vert },1,0,0)\text{, 
}n^{\mu }=\frac{1}{\sqrt{2}}(\frac{1}{\left\vert N\right\vert },-1,0,0)\text{%
, }m^{\mu }=(0,0,m^{a})\text{.}  \label{tet}
\end{equation}%
Eq. (\ref{u}) can be rewritten in the form of a local Lorentz boost along
the x-direction: $u^{\mu }=u^{\mu (0)}\cosh s+a^{\mu (0)}\sinh s$, $a^{\mu
}=a^{\mu (0)}\cosh s+u^{\mu (0)}\sinh s$, where "$^{(0)}$" refers to the
static observer and $a^{\mu (0)}=(0,0,1,0)$, $m^{\mu }=\frac{a^{\mu
}+ib^{\mu }}{\sqrt{2}}$, $b^{\mu }=(0,0,0,\frac{1}{r})$. Then $l^{\mu }=%
\frac{1}{\sqrt{2}}(\frac{\cosh s}{\left\vert N\right\vert },1,\sinh s,0)$, $%
n^{\mu }=\frac{1}{\sqrt{2}}(\frac{\cosh s}{\left\vert N\right\vert }%
,-1,\sinh s,0)$, $m^{\mu }=\frac{1}{\sqrt{2}}(\frac{\sinh s}{\left\vert
N\right\vert },0,\cosh s,\frac{i}{r})$.

For our metric non-vanishing components of the Weyl tensor 
\begin{equation}
C_{\alpha \beta \gamma \delta }=R_{\alpha \beta \gamma \delta }-R_{\gamma
\lbrack \alpha }g_{\beta ]\delta }+R_{\delta \lbrack \alpha }g_{\beta
]\gamma }+\frac{R}{3}g_{\gamma \lbrack \alpha }g_{\beta ]\delta }
\end{equation}

read%
\begin{equation}
C_{0b0d}=N^{2}\frac{B}{12}\gamma _{bd}\text{, }C_{1b1d}=-\frac{B}{12}\gamma
_{bd}\text{, }B=R_{\parallel }+2\lambda \text{, }  \label{d}
\end{equation}%
\begin{equation}
C_{abcd}=\frac{B}{6}(\gamma _{ac}\gamma _{bd}-\gamma _{ad}\gamma _{bc})\text{%
, }C_{0101}=-\frac{N^{2}}{6}B\text{.}
\end{equation}%
Using eq. (\ref{pr}), one can rewrite the value of $B$ in the form%
\begin{equation}
B=16\pi (\rho +p_{\perp })\text{.}  \label{d2}
\end{equation}

Then straightforward calculation gives us%
\begin{equation}
\psi _{0}=-\frac{B\sinh ^{2}s}{8}=\psi _{4}\text{, }\psi _{1}=-\frac{B\cosh
s\sinh s}{8}=\psi _{3}\text{, }\psi _{2}=-\frac{B}{12}(1+\frac{3}{2}\sinh
^{2}s)\text{.}  \label{psi}
\end{equation}%
Determination of the Petrov type is based on studying curvature invariants $%
I,J$ and coefficients $K,L,N$, in certain covariants (see chapter 9.3 in 
\cite{kramer}), constructed from Weyl scalars:%
\begin{equation}
I\equiv \psi _{0}\psi _{4}-4\psi _{1}\psi _{3}+3\psi _{2}^{2}\text{,}
\end{equation}%
\begin{equation}
J=\det \left( 
\begin{array}{ccc}
\psi _{4} & \psi _{3} & \psi _{2} \\ 
\psi _{3} & \psi _{2} & \psi _{1} \\ 
\psi _{2} & \psi _{1} & \psi _{0}%
\end{array}%
\right) \text{,}
\end{equation}%
\begin{equation}
K=\psi _{1}\psi _{4}^{2}-3\psi _{2}\psi _{3}\psi _{4}+2\psi _{3}^{3}\text{, }%
L=\psi _{2}\psi _{4}-\psi _{3}^{2}\text{, }N=12L^{2}-\psi _{4}^{2}I\text{.}
\end{equation}%
One can find that in our case $I=\frac{B^{2}}{48}$, 
\begin{equation}
I^{3}=27J^{2}\text{,}
\end{equation}%
\begin{equation}
N=K=0\text{.}
\end{equation}%
According to the general scheme of classification \cite{kramer}, it means
that there are only two possible cases. 1) $B\neq 0$. Then $\psi _{2}\neq 0$%
, $I\neq 0$ and we have type D, 2) $B=0=I=\psi _{2}$, then we have type O.

It is easy to obtain the condition when the type is O for the entire
manifold. We saw that for the non-compact case $\lambda >0$, so $B$ cannot
vanish since $R_{\parallel }>0$ because of the condition $\rho >0$. It
remains to consider the compact case only. In doing so, we must restrict
ourselves to the presence of the horizon since, as we saw, otherwise $%
\lambda >0$ on the part of manifold, so $B>0$ there in contradiction with
the assumption$.$Using (\ref{rb}), (\ref{pl}) we have $N^{\prime }N^{\prime
\prime }=N^{\prime \prime \prime }N$. As $r\neq 0$, the function $N$ is not
identically constant. Direct integration with the condition $N=0=N^{\prime
\prime }$ on the horizon gives us%
\begin{equation}
N^{\prime \prime }=\pm c^{2}N\text{, }c=const\text{.}
\end{equation}%
With the "+" sign we have, taking into account the boundary condition (\ref%
{01}) at $x=0$, that $r\sim \sinh cx.$ However, eq. (\ref{pr}) then gives us 
$\rho <0$ in contradiction with the assumption $\rho >0$. Hence, we must
choose the sign "-" only. Then, taking into account (\ref{01}) we find $%
r=N^{\prime }=c^{-1}\sin cx$ that corresponds to a sphere. Thus, type O is
possible for a sphere only. Otherwise, Petrov type is D.

Meanwhile, even for a sphere type D is possible as well if the horizon is
absent. Indeed, for a sphere $R_{\parallel }=\frac{2}{r_{0}^{2}}$, whence it
follows from (\ref{spp}) that $B=2r_{0}^{-2}\frac{a}{a-\cos \theta }$. In
the subcase a) discussed above $a\neq 0$, so $B\neq 0$ and the type is D. In
the subcase b) $a=0$ and $B=0$, so we have type O.

We obtained that the gravitational field can be of type O everywhere only
for a sphere. However, if we allow the type to change within the manifold,
one cannot exclude combinations of both types in different regions. Then in
the "normal" region $\lambda >0$, $B>0$ and only type D is possible. In the
"phantom" region where $\lambda <0$, in principle both cases D and O are
possible.

\subsection{Truly naked horizons}

It is worth paying attention to the asymptotic behavior of Weyl scalar near
the horizon, where $N\rightarrow 0$, $r^{\prime }\rightarrow 0$. The
quantity $B\sim \rho +p_{\perp }\rightarrow 0$ near the non-extremal horizon 
\cite{med}. However, it follows from the definition of $s$ (\ref{u}) that
the factors $\cosh s$ and $\sinh s$ diverge like $N^{-1}$. As a result, all $%
\psi ^{\prime }s$ (\ref{psi}) behave like $\frac{B}{N^{2}}$ and we have the
competition of two factors. Consider, for example, the case of the linear
equation of state (\ref{ar}) with $\alpha <1$. This constraint on the value
of $\alpha $ ensures, as we showed, the regularity of the horizon. Then it
follows from (\ref{rb}), (\ref{l}) and (\ref{rn}) that near the horizon $%
\frac{B}{N^{2}}\sim (r_{0}^{2}-r^{2})^{\varepsilon }$, where $\varepsilon =%
\frac{1-3\alpha }{1+\alpha }$. Thus, if we take $\frac{1}{3}<\alpha <1$, all 
$\psi ^{\prime }s$ diverge on the horizon but the geometry remains regular
there.

It was observed recently \cite{vojta} that in spacetimes with static Killing
horizons it may happen that in the free-falling frame some components of the
Riemann and Weyl tensor diverge on the horizon although the Kretschmann
invariant is finite. Such object were called "truly naked black holes"
(TNBH) to distinguish them form just "naked black holes" (NBH) considered in 
\cite{nk1}, \cite{nk2} where some components of the Riemann tensor were
enhanced on the horizon but remained finite. The difficulty in finding
explicit examples of TNBH consists in that the corresponding metrics cannot
be spherically symmetric whereas this is possible for NBH. In this respect,
in the present work we have managed to find "truly naked horizons"
explicitly with the reservation that, as we saw, now they are cosmological.
It is worth reminding that the equation of state $p=-\alpha \rho $ with $%
\alpha >\frac{1}{3}$ is nothing else than the so called dark energy. In our
case, due to the anisotropy, the matter is dark (but not phantom) with
respect to $p_{\perp }$ since $p_{\perp }+\rho >0$ and phantom with respect
to n-direction ($p_{\parallel }+\rho <0$).

\subsection{Case 1}

Now let us discuss briefly case 1 that is more easy. It follows directly
from eqs. (10), (27), (30), (32) of Ref. \cite{vojta} that in the static
frame with the choice of tetrads (\ref{tet}) with $K_{ab}=0$ and $N_{;a}=0$
Weyl scalars $\psi _{0}=\psi _{1}=\psi _{3}=\psi _{4}=0$, while 
\begin{equation}
\psi _{2}=-\frac{R_{\parallel }+R_{\perp }}{12}\text{.}  \label{c}
\end{equation}%
Taking into account (\ref{Nn}), it can be rewritten as $\psi _{2}=-\frac{C}{%
12}$, $C=R_{\parallel }-2\lambda $. It is worth noting that, according to (%
\ref{d}), (\ref{c}), $\lambda $ enters $C$ and $B$ with the opposite signs.

When we are making a local boost along $n$ direction to pass from the static
frame to the free falling one, the Weyl scalars transform in the standard
way according to $\psi _{0}\rightarrow z^{2}\psi _{0}$, $\psi
_{1}\rightarrow z\psi _{1}$, $\psi _{2}\rightarrow \psi _{2}$, $\psi
_{3}\rightarrow z^{-1}\psi _{3}$, $\psi _{4}\rightarrow z^{-2}\psi _{4}$,
where $z=\exp (-s)$. Therefore, in the case under discussion all $\psi
^{\prime }s$, possibly except $\psi _{2}$, vanish in any frame. Thus, there
are no TNBH horizons since none of $\psi ^{\prime }s$ diverges. Petrov Type
is D, if $C\neq 0$ and O, if $C=0$.

\section{Relationship between 1+1, 2+1 and 3+1 theories}

The form of the metric (\ref{4d}) means that, actually, our 4D system is
reduced to 2+1 gravity. For the particular cases such relationship was
discussed earlier. It was observed that one can generate 3+1 cosmic strings
from 2+1 point particles source (see \cite{bar} and references therein). In
a similar way, one may generate a black string from the BTZ black hole \cite%
{kalop}, \cite{btz1}, \cite{btz2} (in our case 3+1 black string is generated
by a cosmological horizon), provided the 3+1 stress-energy tensor has a
certain form \cite{lz1}, \cite{lz2}. In doing so, one obtains a cylindrical
or toroidal 3+1 system depending on whether the variable $n$ changes in an
infinite range or points $n$ and $n+L$ ($L$ is some constant) should be
identified. However, it is worth stressing that in our approach the line
element (\ref{4d}) was not taken \textit{a priori}. It followed from (i) the
condition $K_{ab}=0$, (ii) the (na) Einstein equations, (iii) the assumption
that $N$ does not depend on $n$. In doing so, the stress-energy tensor (\ref%
{tmn}) is anisotropic in the 3+1 world but its reduced form in the 2+1
subspace corresponds to the perfect fluid with the pressure $p_{\perp }$. If 
$p_{\perp }<0$, this implies that the 2+1 version of the strong energy
condition is violated \cite{bar} and so is its 3+1 counterpart because of (%
\ref{pp}). However, the situation with NEC may be quite different for the
3+1 and 2+1 versions. For example, if $p_{\perp }<0$ but $p_{\perp }+\rho >0$%
, the 2+1 version of NEC is satisfied but its 3+1 counterpart is certainly
violated due to (\ref{pp}). In other words, 2+1 tension stars \cite{cf2}
generate in our case 3+1 phantom matter.

As far as classification of gravitational field is concerned, it is worth
noting that although the Weyl tensor vanishes in 3D spacetimes, the fact
that our 2+1 metric is obtained from the 3+1 one enabled us to implement the
standard Petrov classification directly on the basis of 4D Weyl tensor.

The above consideration is generalized easily to the case of the non-zero
cosmological constant $\Lambda $. Then we must replace in formulas $p_{\perp
}\rightarrow p_{\perp }^{(tot)}=p_{\perp }^{(m)}+p^{(\Lambda )}$, $\rho
\rightarrow \rho ^{(tot)}=\rho ^{(m)}+\rho ^{(\Lambda )}$, where $\rho
^{(m)} $ and $p^{(m)}$ is the contribution of matter, $\rho ^{(\Lambda )}=%
\frac{\Lambda }{8\pi }$, $p^{(\Lambda )}=-\frac{\Lambda }{8\pi }$. Then eq. (%
\ref{pp}) is replaced by $\rho ^{(tot)}+p_{\parallel }^{(tot)}=\rho
^{(m)}+p_{\parallel }^{(m)}=2p_{\perp }^{(tot)}$. If $-\rho ^{(m)}<p_{\perp
}^{(m)}<-p^{(\Lambda )}$, again the 2+1 version of NEC is satisfied but the
3+1 one is not. The connection established in previous sections (where we
assumed $\Lambda =0$) between the sign of the pressure $p_{\perp }$ and the
properties of the topology (including the presence of the horizon) is valid
if $\rho _{tot}>0$. When $\rho _{tot}<0$, the situation may change. For
instance, if there is no matter at all and $\Lambda =-\nu ^{2}<0$, one can
obtain known 2+1 BTZ black holes \cite{btz1}, \cite{btz2} for which $%
p_{tot}>0$, $\rho _{tot}<0$, the r-$\phi $ manifold is open. Then it follows
that $p_{\parallel }^{(tot)}=\frac{3\nu ^{2}}{8\pi }$ in accordance with 
\cite{lz2}.

Properties of matter distribution in 2+1 spacetime without horizons (stellar
configurations) were studied in \cite{cf1}, \cite{cf2}, \cite{cz} on the
basis of the line element written in somewhat different form: 
\begin{equation}
ds^{2}=-N^{2}dt^{2}+\frac{dr^{2}}{V}+r^{2}d\phi ^{2}\text{, }  \label{nr}
\end{equation}%
where $V=f^{-1}=\left( \frac{dr}{dx}\right) ^{2}$. To establish
correspondence between (\ref{nr}) and (\ref{f}) with $n=const$, one can
observe that it follows from (\ref{rro}) that 
\begin{equation}
V=C-\frac{k}{\pi }m(r)\text{, }k=8\pi \text{,}  \label{v}
\end{equation}%
\begin{equation}
m=2\pi \int_{0}^{r}dr\rho _{tot}r\text{,}  \label{mr}
\end{equation}%
eqs. (\ref{v}) and (\ref{mr}) correspond to eq. (4) of \cite{cz} (with $m$
instead of $\mu $). Here $C$ is a constant. (If we require the manifold to
contain no conical singularities at the origin, $C=1$.) In a similar way,
one can rewrite the expression for the pressure (\ref{l}) (analogue of eq.
(3) of \cite{cz}) in the form 
\begin{equation}
kp_{\perp }=\frac{\sqrt{V}}{N}\text{, }N=\int^{r}\frac{drr}{\sqrt{V}}\text{.}
\end{equation}

We would like to point out some differences between Refs. \cite{cf1}, \cite%
{cf2}, \cite{cz} and our paper: (i) we include into consideration both
non-compact and compact r-$\phi $ subspaces, (ii) the matter source turns
out to be phantom in general while in the aforementioned papers it is
assumed that $p>0$, (iii) they deal with stellar configurations (no horizon
is present) while we discuss configurations both with and without horizons.

2+1 metrics can be related not only to 3+1 spacetimes, but also to 1+1 ones.
Such a kind of reduction has been carried out in \cite{ana} for the negative
cosmological constant from BTZ-like solutions to the effective 2D theory
(there are also other examples of similar reduction, in particular, for the
Chern-Simons term \cite{dan}, \cite{gur}). In our case, however, the
presence of matter is essential. If the line element (\ref{4d}) is taken
from the very beginning, it is instructive to perform the reduction to 2D
theory from 4D one directly. By substitution of (\ref{4d}) into the
Einstein-Hilbert action $S_{g}=\frac{1}{16\pi }\int d^{4}x\sqrt{-g}R$,
dropping total derivatives and boundary terms we obtain after integration
over $\phi $ and $n$:%
\begin{equation}
S_{g}=\frac{c}{8}\int d^{2}x\sqrt{\gamma }R_{\parallel }N\text{.}  \label{sg}
\end{equation}%
(where $c=\int dn$), while the matter action 
\begin{equation}
S_{m}=2\pi c\int d^{2}x\sqrt{\gamma }NL\text{,}  \label{sm}
\end{equation}%
where $L$ is the matter Lagrangian. By definition of the stress-energy
tensor, we have the 4D formula $\delta S_{m}=-\frac{1}{2}\int d^{4}x\sqrt{-g}%
T_{\alpha \beta }\delta g^{\alpha \beta }$. The variation of $S_{g}$ with
respect to $\gamma _{ab}$ is calculated according to%
\begin{equation}
\delta \int d^{2}x\sqrt{\gamma }R_{\parallel }N=\int d^{2}x\sqrt{\gamma }%
G_{ab}^{eff}\delta \gamma ^{ab}\text{,}
\end{equation}%
where $G_{ab}^{ef}=\gamma _{ab}\Delta _{2}N-N_{;a;b}$. We should also take
into account that in the case under discussion $T_{\alpha \beta }$ is given
by eq. (\ref{tmn}). Then one obtains easily that the variation of the total
action with respect to $\gamma _{ab}$ gives us eqs. (\ref{nab}), while the
variation with respect to $N$ gives eq. (\ref{pr}).

Thus, the effective 2D action of our theory is represented by the sum of two
parts. The action (\ref{sg}) is the analogue of the gravitation-dilaton
action with the "dilaton" $N$, while the matter action is given by (\ref{sm}%
). In contrast to standard string-inspired 2D dilaton gravity models like 
\cite{cg}, now (\ref{sg}) contains neither kinetic nor potential term. Apart
from this, now the 2D theory is not pure gravitation-dilaton one since it
has non-trivial contents only with the matter contribution (\ref{sm}) taken
into account.

One can try to exploit another kind of 2+1+1 decomposition for the metric (%
\ref{4d}) but the result is similar. Rewriting the metric in the form $%
ds^{2}=h_{ij}dx^{i}dx^{j}+dn^{2}+r^{2}(x)d\phi ^{2}$ with the static 2D
metric $h_{ij}$ ($i,j=0,1$) one can choose $r$ as a dilaton instead of $N$.
Then the curvature $R=R_{2}-2\frac{\Delta _{2}r}{r}$, where now $R_{2}$ and $%
\Delta _{2}$ refer to the metric $h_{ij}$. In the coordinates (\ref{4d}) $%
R_{2}=-2N^{-1}\frac{d^{2}N}{dx^{2}}$. Carrying out integration over $n$ and $%
\phi $ and omitting total derivatives, one again arrives at the
gravitation-dilaton effective Lagrangian which has the form (\ref{sg}) with $%
R_{\parallel }$ replaced by $R_{2}$ and $N$ replaced by $r$. It is
straightforward to check that variation with respect to $h_{ij}$ reproduces
eqs. (\ref{rro}) and (\ref{l}).  

\section{General case}

In this section we discuss case 3, when the lapse function (\ref{nxa})
depends both on $n$ and $x^{a}$. Repeating all computations (we again assume
for simplicity that $\Lambda =0$), we obtain that eqs. (\ref{pr}), (\ref{22}%
) and (\ref{nab}) still hold but now eq. (\ref{pl}) changes to 
\begin{equation}
\lambda =\frac{1}{2}\frac{\Delta _{2}N}{N}=8\pi p_{\perp }-\frac{1}{N}\frac{%
d^{2}N}{dn^{2}}\text{,}  \label{g4}
\end{equation}%
so that 
\begin{equation}
N_{2;a;b}=\frac{\Delta _{2}N}{2}\gamma _{ab}\text{,}  \label{n2}
\end{equation}%
whence, again, it follows that the 2D metric takes the form (\ref{mb}) with $%
N_{2}=N_{2}(x)$ having the same form (\ref{n}). It is seen from (\ref{g4})
that%
\begin{equation}
\frac{\Delta _{2}N}{2}+\frac{d^{2}N_{1}}{dn^{2}}=8\pi p_{\perp
}[N_{1}(n)+N_{2}(x)]\text{.}  \label{xn}
\end{equation}

Consider, as an example, the case of a constant pressure $p_{\perp }$. Then,
isolating in eq. (\ref{xn}) parts depending on $x$ and $n$, we obtain that $%
\frac{d^{2}N_{1}}{dn^{2}}-8\pi p_{\perp }N_{1}=C$ and $\frac{\Delta _{2}N_{2}%
}{2}-8\pi p_{\perp }N_{2}=-C$, where $C$ is a constant, whence $N_{1}=-\frac{%
C}{8\pi p_{\perp }}+M_{1}(n)$, $N_{2}=\frac{C}{8\pi p_{\perp }}+M_{2}(x)$.
Here $M_{1}(n)$ is the solution of eq. (\ref{Nn}) and $M_{2}(x)$ is the
solution of (\ref{pl}). Let $8\pi \rho =\kappa ^{2}=const$, then, by analogy
with (\ref{st}), we find the solution with $p_{\perp }=-\rho $ and,%
\begin{equation}
r=\frac{\sin \kappa x}{\kappa }\text{, }N=A_{1}\cos \kappa n-A_{2}\cos
\kappa x\text{,}
\end{equation}%
where $A_{1}$ and $A_{2}$ are constants. If $A_{1}A_{2}\neq 0$, the lapse
function $N=0$ at $n$, $x$ such that $A_{1}\cos \kappa n-A_{2}\cos \kappa
x=0 $. However, as it obeys neither (\ref{N}) nor (\ref{X}), (\ref{N2}) the
Kretschmann invariant diverges and we have an isotropic singularity instead
of a regular horizon. If $A_{1}=0$ we return to case 2 and if $A_{2}=0$ we
return to case 1.

It is easy to see that now a regular horizon is impossible in general.
Indeed, the finiteness of the Kretschmann invariant requires the finiteness
of quantities $\frac{N_{i}\mid _{j}}{N}$ on the horizon \cite{israel}, \cite%
{med}. For our metric it entails the finiteness of $N^{-1}\frac{d^{2}N_{2}}{%
dx^{2}}=\frac{r^{\prime }(x)}{N}$, $N^{-1}\frac{d^{2}N_{1}}{dn^{2}}$ in the
limit $N\rightarrow 0$, where $N=N_{1}(n)+N_{2}(x)$. Obviously, this cannot
be achieved simultaneously, if both derivatives $\frac{d^{2}N_{2}}{dx^{2}}$
and $\frac{d^{2}N_{1}}{dn^{2}}$ do not vanish identically. One can also see
that solutions with equation of state $p_{\perp }=p_{\perp }(\rho )$ are
impossible at all, if $\rho $ is not constant. Indeed, it is seen from eq. (%
\ref{pr}) that in this case $\rho =\rho (x)$, $p_{\perp }=p_{\perp }(x)$
that is inconsistent with eq. (\ref{xn}).

With a non-zero $\Lambda $, one can also obtain for constant density and
pressure regular solutions without horizons:%
\begin{equation}
N=A_{1}\cosh \kappa n+A_{2}\cosh \kappa x\text{, }r=\frac{\sinh \kappa x}{%
\kappa }\text{,}
\end{equation}

$8\pi \rho _{tot}=-\kappa ^{2}$, $p_{\perp }^{(m)}=-\rho ^{(m)}$, $\rho
^{(m)}=-\frac{\kappa ^{2}+\Lambda }{8\pi }$. For the positivity of $\rho
^{(m)}$ it is necessary that $\Lambda <0$ and $\left\vert \Lambda
\right\vert >\kappa ^{2}$.

\section{Summary and conclusion}

We considered extrinsically flat static spacetimes, with the two-dimensional
metric $\gamma _{ab}(x^{2},x^{3})$ not depending on a spatial coordinate $n$%
, where $n$ and $x^{2},x^{3}$ being Gauss normal coordinates. It turned out
that the spacetimes under discussion are naturally separated to two classes
depending whether the gradient of the lapse function points in the direction
of foliation or orthogonally to it. If the cosmological constant $\Lambda =0$%
, the condition $\rho >0$ leads to positivity of the curvature $R_{\parallel
}$ in $x^{2},x^{3}$manifold for both classes, so that the manifold should be
convex. The first class represents direct generalization of BR-like
spacetimes to the non-spherical metric $\gamma _{ab}$, with the lapse
function $N=N(n)$, $\rho $ and $p_{\parallel }$ depending on coordinates $%
x^{2},x^{3}$ and $p_{\perp }$ depending on $n$.

The second class consists of spacetimes for which $N$ does not depend on $n$%
, so that the original 3+1 system reduces to the direct product of the 2+1
manifold and n-axis, $\frac{\partial }{\partial n}$ being the Killing vector
for the 3+1 metric. It turned out that in this case there exists also an
additional Killing vector $\frac{\partial }{\partial \phi }$ commuting with $%
\frac{\partial }{\partial n}$, $\phi $ having the meaning of the angular
coordinate. For compact 2D $x-\phi $ manifolds ($x=x^{2}$, $\phi =x^{3}$)
from this class the null energy condition is necessarily violated either on
the part of the surface $n=const$ (if there is no horizon) or on the entire
surface (if there is a horizon). The horizon is a cosmological one. For a
linear and Chaplygin gas equation of state we obtained exact explicit
solutions. Their 3+1 source is necessarily violates the null energy
condition, although from the pure 2+1 viewpoint the null energy condition
may be satisfied. In the non-compact case null energy condition is respected
and, moreover, the transversal pressure $p_{\perp }>0$. Thus, there is an
intimate connection between the sign of $p_{\perp }$ and topology of the
section $t=const$, $n=const$.

There exists also the general class 3 for which the lapse function depends
on both $x$ and $n$. This leads to singular "horizons" but in the absence of
a horizon the solution may be regular, as it is seen from the particular
example with the uniform density and pressure.

It turned out that within class 2 the horizon may be "truly naked" in the
sense that some Weyl scalars diverge on it, although the Kretschmann scalar
is finite that gives an explicit example of objects discussed in \cite{vojta}%
. No such objects exist in case 1. Petrov type in case 1 is O, if $%
R_{\parallel }=2\lambda $ and D if $R_{\parallel }\neq 2\lambda $. For case
2 the "naked" character of the horizon does not prevent implementation of
Petrov classification since divergencies completely cancel out each other in
the invariants that enter the scheme of classification. In case 2 also only
types D or O are possible, the corresponding conditions read $R_{\parallel
}=-2\lambda $ and $R_{\parallel }\neq -2\lambda $, respectively. Type O for
the entire manifold is possible for a sphere only, which, however, admits
also the solution of type D everywhere. If the horizon is absent, the Petrov
type is D on the part of manifold where $p_{\perp }>0$. The results are
generalized to the case of non-zero $\Lambda $. If $\Lambda <0$ and $\rho
_{tot}<0$, the curvature $R_{\parallel }<0$, so the 2D manifold becomes
concave.

It is of interest to extend the present results to stationary spacetimes
since it would give possibility to take into account rotation in issues
discussed in Introduction.

\end{document}